\documentclass[preprintnumbers,amsmath,amssymb,twocolumn]{revtex4}

\usepackage{graphicx}% Include figure files
\usepackage{epsfig}
\usepackage{epsf}
\usepackage{verbatim}
\usepackage{multirow}

\begin{document}

\title{ 
Inflationary Perturbations in Palatini Generalised Gravity}

\author{ Nicola Tamanini}
\altaffiliation{\it Department of Physics, University of Trento, 38123 Trento, Italy}
\author{Carlo R.~Contaldi}
\affiliation{\it Theoretical Physics, Blackett Laboratory, Imperial
  College, London, SW7 2BZ, UK}
\email{c.contaldi@imperial.ac.uk}

\begin{abstract}
  We examine the generation of primordial perturbations during an
  inflationary epoch in generalised theories of gravity when the
  equations of motion are derived using the Palatini variational
  principle. Both $f(R)$ and Scalar-Tensor theories are considered and
  we compare our results with those obtained under the conventional
  metric formalism. Non-linear generalisations of the action lead to
  different theories under the two variational choices and we obtain
  distinct results for scalar and tensor spectral indices and their
  ratio. We find the following general result; inflation driven solely
  by $f(R)$ modifications alone do not result in suitable curvature
  perturbations whilst Scalar-Tensor theories generate nearly scalar
  invariant curvature perturbations but no tensor modes.
\end{abstract}

\maketitle

\section{Introduction}

Today, more than ever, there is an element of doubt underlying the
confidence with which we claim to understand the nature and behaviour
of the universe. The so-called standard $\Lambda$ Cold Dark Matter
($\Lambda$CDM) model produces extraordinarily good fits to numerous
observations in the era of precision cosmology we live in. However,
the theory of General Relativity (GR) underpinning the dynamical
workings of the standard model requires a number of supporting
partners throughout the entire history of the universe to bring about
this remarkable agreement with data. A scalar field is needed to drive
the inflationary period after the Big Bang; a strange form of energy,
called Dark Energy, is required for the late time observed
acceleration; while an unknown kind of pressureless matter, called
Dark Matter, is postulated to account for the missing mass of galaxies
and galaxy clusters. As such, beyond solar system scales, where the
theory has been well tested and confirmed, GR always requires the
introduction of some exotic form of energy-density to stand up against
confrontation with observations. This uncomfortable situation has led
to many attempts at modifying GR itself in the search of a generalised
theory that can explain the observed universe without requiring the
addition of dark companions.  Although none of the modifications seem
to have the general and unifying character of GR itself, it can be
argued that some models can address specific problems in a more
natural way.

In this work we deal with a class of theories which has been denoted
in the literature under the name of \textit{Generalised} or
\textit{Extended Gravity} \cite{Hwang:1990re,Capozziello:2007ec}. The
starting point of these theories is a general modification of the
Einstein-Hilbert action consisting both in the introduction of a
scalar field $\phi$ and the addition of terms which are non-linear in
the curvature scalar $R$. The general form such an action can assume
is given in Section~\ref{PalatiniFormulation}. This class of
gravitational theories incorporates both sub-classes given by $f(R)$
theories \cite{Sotiriou:2008rp,DeFelice:2010aj} and Scalar-Tensor
theories \cite{Fujii:2003pa,Faraoni:2004pi}. The first treats only
non-linear terms in $R$ given by an arbitrary function $f(R)$, without
introducing an extra scalar field. The second considers instead all
the possible modifications given by $\phi$, but $R$ appears only
linearly in the action (though its coefficient is a general function
of $\phi$). Further constrains on the form of $f$ or on the scalar
field terms can lead to specific theories which have been used
extensively in literature trying to solve many different problems. For
example, some of the very first models of inflation were an $f(R)$
theory based on an $R^2$ term \cite{Starobinsky:1980te} and a
minimally coupled Scalar-Tensor theory \cite{Guth:1980zm}. The
well-known Brans-Dicke theory \cite{Brans:1961sx} can also be regarded
as the first Scalar-Tensor theory of gravitation. More recently $f(R)$
theories have been invoked as a possible explanation for the late time
acceleration of the universe \cite{Sotiriou:2008rp,DeFelice:2010aj}
without the need to introduce Dark Energy. On the other hand
Scalar-Tensor theories seems to be even more promising as models for
inflation \cite{GarciaBellido:1995fz,Barrow:1995fj}.  Nevertheless
there have been several attempts using $f(R)$ and Scalar-Tensor
theories to characterize both inflation and late time acceleration,
respectively
\cite{Kaneda:2010qv,DeFelice:2010aj,Gannouji:2006jm}. Some models go
even further in trying to solve the dynamical riddle of early-time inflation
and late-time acceleration in a unified manner
\cite{Sotiriou:2005hu,Nojiri:2005pu,Nojiri:2007as}.

Once a modified action has been chosen there is a further degree of
freedom when generalising gravity. The variational principle used to
obtain the field equations for the model must be chosen. There are two
main variational methods widely used in the literature
\cite{Misner:1974qy} (see also \cite{Ferraris:1982} for a historical
perspective on their origin). The first is the usual metric
variational principle used by Hilbert in the first derivation of the
Einstein equation with a variational method. The metric formalism
obtains the gravitational field equations by varying the action only
with respect to the space-time metric tensor $g_{\mu\nu}$, which is
assumed to be symmetric. An alternative is known as the Palatini
variational principle and consists in varying the action with respect
to the metric and the torsionless connection
$\hat\Gamma^\lambda_{\mu\nu}$ independently. Here the independent
connection $\hat\Gamma^\lambda_{\mu\nu}$ is not assumed to be, a
priori, the more commonly used Levi-Civita connection
$\Gamma^\lambda_{\mu\nu}$ for which the additional metric condition
$\nabla_\mu g_{\alpha\beta}=0$ holds. In the Palatini formalism this
condition is in fact obtained dynamically by varying the action rather
than being imposed axiomatically from the beginning. Crucially, if the
action is non-linear in the curvature scalar, the Palatini method of
variation leads to a different theory than the metric variation
\cite{Sotiriou:2008rp}. This means that the choice of variational
method is not just a question of formalism, but can in principle lead
to different physical outcomes. Both methods have their own advantages
and handicaps, but the metric variational principle has been on the
whole more utilized in the literature. Because of this, many issues in
generalised gravity have already been studied carefully in the metric
variation formalism but have not been looked at in depth in the
context of the Palatini formalism. The aim of this work is to deal
with precisely one of these issues, namely the characterisation of
super-horizon, primordial perturbations generated during an epoch of
inflation. These have been examined extensively in the metric case
\cite{Hwang:1990re,Hwang:1996xh,DeFelice:2010aj}, but the Palatini
approach has yet to be studied in detail. 

The standard theory of inflation is basically a Scalar-Tensor
theory where a potential dominated, minimally coupled scalar field
drives inflation \cite{Mukhanov:2005sc,Liddle:2000cg}. However some
$f(R)$ models succeed in building theories where the universe inflates without
the introduction of a scalar field. This can be achieved in both
metric \cite{Starobinsky:1980te,DeFelice:2010aj,Kaneda:2010qv} and
Palatini \cite{Wang:2004vs,Sotiriou:2005hu} formalisms. In the latter
case, models giving bouncing cosmologies have been also proposed
recently \cite{Barragan:2009sq}. It is then interesting to ask how the
primordial perturbations have been generated in the various models. To
do this we carry out a general analysis of perturbations for Palatini
generalised gravity in which the specific choice of the model is left
for the end of the calculations. This analysis will include all the
$f(R)$ and Scalar-Tensor models in Palatini formalism, leaving
standard inflation as a particular case. The aim is to determine the
power spectra and spectral indices for scalar and tensor
perturbations.

The motivation for focusing on the characterisation of the primordial
perturbations stems from the observational front-line. Over the next
decade observations of the Cosmic Microwave Background (CMB) will
constrain to unprecedented level the spectral index of primordial
scalar (curvature) perturbations and the amplitude of tensor
(gravitational wave) perturbations. In particular, if detected, the
tensor modes will allow us to infer the energy scale and nature of the
inflaton potential. Conversely, in the context of generalised
theories, it may allow us to constrain the nature of the modifications.

This \textit{paper} is organized as follows. In
Section~\ref{PalatiniFormulation} we provide a short review of the
Palatini approach to generalised gravity, where the gravitational
field equations and the background cosmological dynamics are
derived. Section~\ref{cosmologicalperturb} is dedicated to developing
formalism and dynamics of cosmological perturbations in these
models. The main section of the paper is Section~\ref{pertinfl} where
we examine cosmological perturbations generated during an inflationary
epoch and derive the form of spectral indices for both scalar and
tensor modes and their ratio. In section~\ref{sec:examples} we apply
these results to f(R) gravity and scalar-tensor theories and provide
an example with non-minimal inflation. We conclude by summarising our
main results in Section~\ref{sec:disc}.

Throughout the paper, unless otherwise specified, we work in units
such that $c = 8\pi G = 1$. Greek and Latin indices run from 0 to 3
and 1 to 3, respectively. The Minkowski metric is taken to be
$\eta_{\mu\nu}=\mbox{diag}(-1,+1,+1,+1)$.

\section	{Palatini Formulation of Generalised Gravity}
\label{PalatiniFormulation}

We begin with a review of the Palatini approach to generalised
gravity, where the connection is varied independently from the metric
(see e.g. \cite{Ferraris:1992dx,Sotiriou:2008rp} for more in depth
reviews). Consider a generalised gravitational theory described by the
action
\begin{eqnarray}
S = \int d^4x \sqrt{-g}
&\Biggl[&\frac{1}{2} f\left(\hat{R}(g_{\mu\nu},\hat{\Gamma}^\alpha_{\beta\gamma}),\phi\right) \nonumber\\
&& + \mathcal{L}_\phi(g_{\mu\nu},\phi) + \mathcal{L}_M(g_{\mu\nu},\Xi)\Biggl]\,,
\label{action}
\end{eqnarray}
where \textit{f} is an arbitrary function of $\hat{R}$ and of a
generic scalar field $\phi$ while $\mathcal{L}_M$ is the matter
Lagrangian depending on some matter fields denoted collectively by
$\Xi$. Here, the curvature scalar $\hat{R}=g^{\mu\nu}\hat{R}_{\mu\nu}$ is the contraction of the Ricci tensor defined by the torsionless independent connection $\hat{\Gamma}^\alpha_{\beta\gamma} \,$,
\begin{equation}
\hat{R}_{\mu\nu} = \hat{\Gamma}^\alpha_{\mu\nu,\alpha} - \hat{\Gamma}^\alpha_{\mu\alpha,\nu} + \hat{\Gamma}^\alpha_{\alpha\lambda}\hat{\Gamma}^\lambda_{\mu\nu} - \hat{\Gamma}^\alpha_{\mu\lambda}\hat{\Gamma}^\lambda_{\alpha\nu} \,.
\end{equation}
The Lagrangian for the scalar field $\phi$ is taken to be
\begin{equation}
\mathcal{L}_\phi = - \frac{1}{2}\omega(\phi)(\partial\phi)^2 - V(\phi)\,,
\end{equation}
from which we define the energy-momentum tensor
\begin{equation}
T^{(\phi)}_{\mu\nu} = \omega (\nabla_\mu\phi)(\nabla_\nu\phi) - g_{\mu\nu}[\frac{1}{2}\omega(\partial\phi)^2 + V(\phi)]\,,
\label{Tscalarfield}
\end{equation}
where $\nabla$ is the covariant derivative associated with the usual
Levi-Civita connection $\Gamma^\alpha_{\beta\gamma}$. In general,
quantities formed with the independent ``Palatini'' connection will be
denoted by an \textit{overhat} to distinguish them from quantities
formed with the Levi-Civita connection. For example, the curvature
scalar and Ricci tensor we know from canonical GR are denoted by $R$
and $R_{\mu\nu}$.  Note that in the Palatini variational principle the
matter Lagrangian $\mathcal{L}_M$ does not depend on the independent
connection $\hat{\Gamma}^{\lambda}_{\mu\nu}$. Allowing the Lagrangian
for this dependence leads to another formalism, namely the
metric-affine variational principle
\cite{Sotiriou:2008rp,Capozziello:2009mq}, which we will not consider
in the current work.

\subsection*{Field Equations}

Variation of the action (\ref{action}) with respect to the metric gives
\begin{equation}
F(\hat{R},\phi)\hat{R}_{\mu\nu} - \frac{1}{2}f(\hat{R},\phi)g_{\mu\nu} = T_{\mu\nu} + T^{(\phi)}_{\mu\nu}\,,
\label{fieldeqmet}
\end{equation}
where $T_{\mu\nu}$ is the usual matter energy-momentum tensor and the function $F$ is defined by
\begin{equation}
F:=\frac{\partial f(R,\phi)}{\partial R} \,.
\end{equation}
Varying with respect to the independent connection
$\hat{\Gamma}^{\lambda}_{\mu\nu}$ gives instead the condition
\begin{equation}
\hat{\nabla}_\mu(\sqrt{-g}g^{\alpha\beta}F) = 0 \,.
\label{fieldeqconn}
\end{equation}

Equations for the scalar field are obtained varying
action~(\ref{action}) with respect to $\phi$. However they are not of
interest in what follows and we omit them here. In order to recover
GR we can simply choose $f(\hat{R},\phi)=\hat{R}$
in~(\ref{action}). In this case from (\ref{fieldeqconn}) we find that
the connection coincides with the usual Levi-Civita
$\hat{\Gamma}^{\lambda}_{\mu\nu}\equiv \Gamma^{\lambda}_{\mu\nu}$ and
(\ref{fieldeqmet}) reduces to the standard Einstein field equation.

If we define a new metric $h_{\mu\nu}$ conformally connected to
$g_{\mu\nu}$ by $h_{\mu\nu}=F\,g_{\mu\nu}$, (\ref{fieldeqconn})
becomes
\begin{equation}
\hat{\nabla}_{\mu}(\sqrt{-h}\,h^{\alpha\beta}) = 0\,,
\end{equation}
signifying that $\hat{\Gamma}^{\lambda}_{\mu\nu}$ is the Levi-Civita
connection with respect to the new metric $h_{\mu\nu}$. We then find
\begin{eqnarray}
\hat{\Gamma}^{\lambda}_{\mu\nu} & = & \frac{1}{2} h^{\lambda\sigma} (h_{\nu\sigma,\mu} + h_{\mu\sigma,\nu} + h_{\mu\nu,\sigma}) \nonumber \\
     & = & \Gamma^{\lambda}_{\mu\nu} + \frac{1}{2F} [2\delta^\lambda_{\,\,(\mu}\partial^{\,}_{\nu)}F + g_{\mu\nu}g^{\lambda\sigma}\partial_\sigma F] \,.
\label{palatiniconn}
\end{eqnarray}
It is useful to use (\ref{palatiniconn}) to replace the Palatini
connection wherever it appears in our equations. The hatted curvature
scalar and Ricci tensor then become \cite{Wang:2004vs}
\begin{eqnarray}
\hat{R}_{\mu\nu} =R_{\mu\nu} &+&  \frac{3}{2F^2} (\nabla_\mu F) (\nabla_\nu F) \nonumber\\ &-& \frac{1}{F}\nabla_\mu\nabla_\nu F - \frac{1}{2F}g_{\mu\nu}\nabla_\sigma\nabla^\sigma F\,,
\label{hatricci}
\end{eqnarray}
and
\begin{equation}
\hat{R} = R - \frac{3}{F}\nabla_\mu\nabla^\mu F + \frac{3}{2F^2}(\nabla_\mu F) (\nabla^\mu F) \,.
\end{equation}

We can also write the field equation (\ref{fieldeqmet}) as the
canonical Einstein field equation plus an effective energy-momentum
tensor containing the dependence on $f$ \cite{Koivisto:2005yc}
\begin{equation}
G_{\mu\nu} = T_{\mu\nu} + T^{(\phi)}_{\mu\nu} + T^{(\mbox{eff})}_{\mu\nu}\,,
\label{fieldeqfinal}
\end{equation}
with
\begin{eqnarray}
T^{(\mbox{eff})}_{\mu\nu} & = & (1-F)R_{\mu\nu} - \frac{3}{2F} (\nabla_\mu F) (\nabla_\nu F) + \nabla_\mu\nabla_\nu F \nonumber\\ & + & \frac{1}{2}g_{\mu\nu} \left[(f-R)+(1-\frac{3}{F})\nabla_\sigma\nabla^\sigma F \right.\nonumber\\ && \qquad\qquad\qquad \left.+\frac{3}{2F}(\nabla_\sigma F) (\nabla^\sigma F)\right]\,.
\end{eqnarray}
Since the Palatini connection does not appear explicitly in the final
form of the field equation, it plays the role of an
of auxiliary field. The physically relevant connection, i.e. the one
determining the free-falling motion, is still
$\Gamma^\lambda_{\mu\nu}$. All the energy-momentum tensors in the
right hand side of (\ref{fieldeqfinal}) are indeed covariantly
conserved by $\nabla$ \cite{Koivisto:2005yk}.

\subsection*{Background Cosmological Equations}

Assuming the universe is described by the general
Friedmann-Robertson-Walker (FRW) metric, we can find equations for the
background cosmological evolution. This assumption is completely
independent of the theory of gravity we choose since it relies only on
treating the universe as homogeneous and isotropic.

The (unperturbed) general FRW metric reads
\begin{equation}
ds^2 = -dt^2 + a^2(t) \left[\frac{dr^2}{1-Kr^2}+r^2d\Omega^2\right]\,,
\label{FRWmetric}
\end{equation}
where $a(t)$ is the scale factor of the universe, $K = 0,1,-1$ is
the spatial curvature and $d\Omega^2:=d\theta^2+\sin^2\theta d\varphi^2$. Moreover we take the matter energy-momentum
tensor to be of the form of a perfect fluid
\begin{equation}
T_{\mu\nu} = (\rho_M+p_M)u_\mu u_\nu + p_M\, g_{\mu\nu}\,,
\label{Tperfluid}
\end{equation}
with $u^\mu$ denoting the four-velocity of an observer comoving with
the fluid. $\rho_M$ is the energy-density and $p_M$ is the pressure of
the matter fluid. If we further consider the scalar field $\phi$ to be
homogeneous in space, i.e. depending only on the physical time $t$, we
can rewrite its energy-momentum tensor (\ref{Tscalarfield}) in the
form (\ref{Tperfluid}) defining
\begin{eqnarray}
\rho_\phi = \frac{1}{2}\omega\dot{\phi}^2 + V(\phi) \qquad
p_\phi = \frac{1}{2}\omega\dot{\phi}^2 - V(\phi)\,,
\end{eqnarray}
as its energy-density and pressure, respectively. Here a dot denotes
differentiation with respect to the physical time $t$.

Having made these assumptions we can derive equations for the scale
factor $a(t)$ in the Palatini formalism.  Using the FRW metric
(\ref{FRWmetric}), the non vanishing components of (\ref{hatricci})
are \cite{Wang:2004vs}
\begin{eqnarray}
&&\hat{R}_{00} = -3\frac{\ddot{a}}{a} + \frac{3\dot{F}^2}{2F^2} - \frac{3}{2F}\nabla_0\nabla_0F \,,\\
&&\hat{R}_{ij} = [a\ddot{a} + 2\dot{a}^2 + 2K + \frac{a^2}{2F}\nabla_0\nabla_0F]\delta_{ij} + \frac{\dot{F}}{F}\Gamma^0_{ij} \,. \qquad
\end{eqnarray}
Substituting these two equations back in (\ref{fieldeqmet}), we find
\begin{eqnarray}
6F\left(H^2+\frac{\dot{F}^2}{4F^2}+\frac{H\dot{F}}{2F}+\frac{K}{a^2}\right) - f & = & \rho+3p \label{Friedmann}\,, \\
2F\left(\dot{H}+\frac{K}{a^2}\right) - \frac{3\dot{F}^2}{2F} - H\dot{F} + \ddot{F} & = & -\rho-p \label{acceq}\,, \qquad
\end{eqnarray}
where $\rho=\rho_\phi+\rho_M$ and $p=p_\phi+p_M$. Equation
(\ref{Friedmann}) can be considered as the modified Friedmann
equation, whilst (\ref{acceq}) is the modified acceleration
equation (or second Friedmann equation). The first one governs the evolution
of the scale factor in the cosmology of generalised Palatini gravity
theories \cite{Wang:2004vs}; the second one will turn out useful later
on.

\section{Cosmological Perturbations}
\label{cosmologicalperturb}

In this section we develop the general formalism of cosmological
perturbations, presenting equations for scalar and tensor
perturbations in Palatini generalised gravity (see
\cite{Hwang:1990re,DeFelice:2010aj} for the metric case). 

\subsection*{Preliminaries}

The general perturbed FRW metric is
\begin{eqnarray}
ds^2 &=& -(1+2\alpha)dt^2 - 2a(t)(\partial_i\beta-b_i)\,dt\,dx^i \nonumber\\
&&\quad + a^2(t) (\delta_{ij}+2\psi\delta_{ij} + 2\partial_i\partial_j\gamma \nonumber\\&&\qquad\qquad
 +2\partial_jc_i+h_{ij})\,dx^idx^j\,,
\end{eqnarray}
where $\alpha,\,\beta,\,\psi,\,\gamma$ are scalar perturbations,
$b_i,\,c_i$ are vector perturbations and $h_{ij}$ is the tensor
perturbation. As usual, we have decomposed the independent degrees of
freedom of the metric perturbations according to their transformation
properties under spatial rotations. Because of the assumed background
homogeneity and isotropy, all the physical quantities have to be
invariant under purely spatial gauge transformations. Consequently, we
want to derive gauge invariant equations for the perturbations. In
general there are several gauge invariants we can build, but among
them, we will need only four quantities. The two gauge invariants
\begin{eqnarray}
\chi := a(\beta+a\dot{\gamma})\,, \qquad
 A := 3(H\alpha-\dot\psi)-\frac{\triangle}{a^2}\chi\,,
\end{eqnarray}
where $\triangle$ denotes the three-space Laplacian, will be useful to
simplify the equations later; while the two invariants
\begin{eqnarray}
\mathcal{R}_{\delta\phi} := \psi-\frac{H}{\dot{\phi}}\delta\phi\,, \qquad 
\mathcal{R}_{\delta F} := \psi-\frac{H}{\dot{F}}\delta F\,,
\label{gaugeinvs}
\end{eqnarray} 
will play a role in the discussion of the gauge choice. Here
$\delta\phi$ and $\delta F$ are the perturbations of $\phi$ and $F$,
respectively.

Before writing down equations for the cosmological perturbations we
discuss the choice of a specific gauge. To date, several choices have
been considered in the literature. The most frequent choices are the
Synchronous gauge $\alpha=0$, the Comoving gauge $v=0$ (where $v$ is
the matter fluid velocity perturbation) and the Newtonian gauge
$\chi=0$. A more convenient choice for this work is to choose either
the \textit{Uniform-F gauge} $\delta F=0$ or the \textit{Uniform-field
  gauge} $\delta\phi=0$.

The first gauge is the more useful to work with when we are restricted to
$f(R)$ theories of gravity. In this class of theories all the
contributions of the scalar field $\phi$ (and of its perturbation
$\delta\phi$) are absent. From the definitions (\ref{gaugeinvs}) we
see immediately that setting $\delta F=\delta\phi=0$ gives
$\mathcal{R}_{\delta F}=\mathcal{R}_{\delta\phi}=\psi$. This means
that $\psi$ becomes then a gauge invariant and we can try to find an
equation governing its evolution. 

The Uniform-field gauge is instead used in Scalar-Tensor theories of
gravity. In this class of theories $F$ depends only on $\phi$, i.e. we
have $f(R,\phi)\equiv F(\phi)R$, and we get $\delta
F=F_{,\phi}\delta\phi$. Choosing $\delta\phi=0$ immediately gives
$\delta F=0$. These are the same conditions we had in the $f(R)$
theories with the Uniform-$F$ gauge. All the results obtained setting
$\delta F=\delta\phi=0$ will then be valid for both the classes of
theories. 

Unfortunately this analysis does not hold for more general theories,
where both a scalar field and a non-linear coupling of $R$
appear. Setting $\delta\phi=0$ does not imply $\delta F=0$ since in
that case
we have $\delta F=F_{,\phi}\delta\phi+F_{,R}\delta R$. Another choice
of gauge is probably more appropriate in the general case but we will
restrict ourselves to the pure $f(R)$ or $F(\phi)$ case in this work.

\subsection*{Perturbation Equations}

Since we deal with perturbations during the inflationary phase of the universe, we do not consider any kind of matter (a part from the scalar field $\phi$). From now on we then set $\mathcal{L}_M=\rho_M=p_M=0$. Cosmological perturbation equations in the Palatini context have been derived in \cite{Koivisto:2005yc} (see also \cite{Tsujikawa:2007tg} in which some typos are corrected). In (spatial) Fourier space, the equations governing the \textit{scalar} perturbations read
\begin{widetext}
\begin{eqnarray}
{\cal G}A + \frac{3K-k^2}{a^2}\psi &+&
\frac{1}{2F}\left[\frac{3\dot{F}^2}{2F}+3H\dot{F}-\omega\dot\phi^2\right]\alpha
\nonumber \\ &=& \frac{1}{2F} \left\{ -\omega\dot\phi\dot{\delta\phi}
  - \frac{1}{2}\left[\omega_{,\phi}\dot\phi^2 +
    (2V-f)_{,\phi}\right]\delta\phi \right. \nonumber \\ && \qquad +
\left. \left[3H^2 - \frac{3\dot{F}^2}{4F^2} - \frac{R}{2} +
    \frac{3K^2+k^2}{a^2}\right]\delta F + 3\,{\cal G}\dot{\delta F} \right\}\,, \label{scalar1}
%\left(H+\frac{\dot{F}}{2F}\right)A + \frac{3K-k^2}{a^2}\psi &+& \frac{1}{2F}\left[\frac{3\dot{F}^2}{2F}+3H\dot{F}-\omega\dot\phi^2\right]\alpha \nonumber \\ &=& \frac{1}{2F} \left\{ -\omega\dot\phi\dot{\delta\phi} - \frac{1}{2}\left[\omega_{,\phi}\dot\phi^2 + (2V-f)_{,\phi}\right]\delta\phi \right. \nonumber \\ && \qquad + \left. \left[3H^2 - \frac{3\dot{F}^2}{4F^2} - \frac{R}{2} + \frac{3K^2+k^2}{a^2}\right]\delta F + 3\left(H+\frac{\dot{F}}{2F}\right)\dot{\delta F} \right\}\,, \label{scalar1}
\end{eqnarray}
\begin{eqnarray}
{\cal G}\alpha - \dot\psi + \frac{K}{a^2}\chi
& = & \frac{1}{2F} \left[\omega\dot\phi\delta\phi -
  \left(H+\frac{3\dot{F}}{2F}\right)\delta F + \dot{\delta
    F}\right]\,, \label{scalar2}
%\left(H+\frac{\dot{F}}{2F}\right)\alpha - \dot\psi + \frac{K}{a^2}\chi
%& = & \frac{1}{2F} \left[\omega\dot\phi\delta\phi -
%  \left(H+\frac{3\dot{F}}{2F}\right)\delta F + \dot{\delta
%    F}\right]\,, \label{scalar2}
\end{eqnarray}
\begin{eqnarray}
\dot\chi + \left(H+\frac{\dot{F}}{F}\right)\chi - \alpha - \psi & = & \frac{1}{F}\delta F\,, \label{scalar3}
\end{eqnarray}
\begin{eqnarray}
\dot{A} + \left(H+{\cal G}\right)A +
\frac{3\dot{F}}{2F}\dot\alpha & + & \left[ 3\dot{H} +
  \frac{3H\dot{F}}{2F} + \frac{3\ddot{F}}{F} - \frac{3\dot{F}^2}{F^2}
  + \frac{2}{F}\omega\dot\phi^2 - \frac{k^2}{a^2}\right]\alpha
\nonumber \\ &=& \frac{1}{2F} {\Biggl\{} 4\omega\dot\phi\dot{\delta\phi} + \left[2\omega_{,\phi}\dot\phi^2 + (f-2V)_{,\phi}\right]\delta\phi  \nonumber \\ && \qquad \left. + \left[\frac{3\dot{F}^2}{F^2}+6(H^2+\dot{H})-R+\frac{k^2}{a^2}\right]\delta F + \left(3H-\frac{6\dot{F}}{F}\right)\dot{\delta F} + 3\ddot{\delta F} \right\} \,. \label{scalar4}
%\dot{A} + \left(2H+\frac{\dot{F}}{2F}\right)A +
%\frac{3\dot{F}}{2F}\dot\alpha & + & \left[ 3\dot{H} +
%  \frac{3H\dot{F}}{2F} + \frac{3\ddot{F}}{F} - \frac{3\dot{F}^2}{F^2}
%  + \frac{2}{F}\omega\dot\phi^2 - \frac{k^2}{a^2}\right]\alpha
%\nonumber \\ &=& \frac{1}{2F} {\Biggl\{} 4\omega\dot\phi\dot{\delta\phi} + \left[2\omega_{,\phi}\dot\phi^2 + (f-2V)_{,\phi}\right]\delta\phi  \nonumber \\ && \qquad \left. + \left[\frac{3\dot{F}^2}{F^2}+6(H^2+\dot{H})-R+\frac{k^2}{a^2}\right]\delta F + \left(3H-\frac{6\dot{F}}{F}\right)\dot{\delta F} + 3\ddot{\delta F} \right\} \,. \label{scalar4}
\end{eqnarray}
\end{widetext}
Where we have defined ${\cal G} = H+\dot F/2F$. These equations are obtained by perturbing the $G^0_{\,0}$,
$G^0_{\,i}$, $G^i_{\,j}-\frac{1}{3}\delta^i_{\,j}G^k_{\,k}$ and
$G^k_{\,k}-G^0_{\,0}$ components of the field equations, respectively.

The equations for vector and tensor perturbations in Palatini
formalism are the same of the corresponding ones in the metric
formulation. This is due to the fact that both $R$ and $\phi$ do not
have either vector or tensor components being pure scalars. The equation
for \textit{vector} perturbations is
\begin{equation}
\frac{k^2-2K}{2a^2}\Psi_i = 0\,,
\label{vectoreq}
\end{equation}
where we have introduced the gauge invariant
$\Psi_i=b_i+a\dot{c}_i$. Equation (\ref{vectoreq}) is exactly the same
equation we find in canonical GR (without matter). It tells us that
vector perturbations will not be generated during inflation, making
these modes uninteresting in generalised gravity too.

The gravitational wave equation, the equation for \textit{tensor} perturbations, is given by
\begin{equation}
\ddot{h}_{ij} + \left(3H+\frac{\dot{F}}{F}\right)\dot{h}_{ij} + \frac{k^2+2K}{a^2}h_{ij} = 0 \,.
\label{gweq}
\end{equation}
In this case the equation is modified from the corresponding equation
in canonical GR. The only difference appears in an additional damping
term modulated by the form of $f$.

\section{Perturbations from Inflation}
\label{pertinfl}

We now focus on the main subject of this paper. Cosmological
perturbations generated during the inflationary stage have already
been considered in the metric formulation for generalised gravity
models. Power spectra amplitudes and indices were calculated for both
scalar and tensor modes in \cite{Hwang:1996xh,DeFelice:2010aj}. An
analysis of the growth of matter density perturbations has been
discussed in both metric and Palatini formalisms
\cite{Tsujikawa:2007tg}. However, the calculation of scalar and tensor
power spectra in the Palatini formalism has not been looked at in
detail. In obtaining our results we follow the procedure carried out
in \cite{Hwang:1996xh,DeFelice:2010aj} and find some important
differences in the observational consequences between the two formalisms.

\subsection*{Scalar Perturbations}

Current observations suggest that the universe is spatially flat and
we therefore restrict our calculation to the $K=0$ case. Furthermore
imposing the gauge choice discussed above ($\delta\phi=\delta F=0$)
the right hand sides of equations (\ref{scalar1})-(\ref{scalar4})
vanish and from (\ref{scalar2}) we then obtain
\begin{equation}
\alpha=\frac{\dot\psi}{\cal G}\,,
\label{alpha}
\end{equation}
%where, in order to semplify the equations, we have defined the function
%\begin{equation}
%S=\left(H+\frac{\dot{F}}{2F}\right)\,.
%\end{equation}
Inserting (\ref{alpha}) into (\ref{scalar1}) gives
\begin{equation}
A = \frac{1}{S} \left[\frac{k^2}{a^2}\psi - \frac{\dot\psi}{2FS} \left(\frac{3\dot{F}^2}{2F}+3H\dot{F}-\omega\dot\phi^2\right)\right] \,.
\label{A}
\end{equation}
Putting (\ref{alpha}) and (\ref{A}) into (\ref{scalar3}) and using the
background equation (\ref{acceq}), we obtain a second-order
differential equation for the gauge invariant perturbation $\psi$
\begin{equation}
\ddot\psi + \left(3H+\frac{\dot{Q}}{Q}\right)\dot\psi + \frac{k^2}{a^2}\psi = 0\,,
\label{eqpsi}
\end{equation}
where we have defined
\begin{equation}
Q := \frac{\omega\dot\phi^2}{S^2} \,.
\label{defQ}
\end{equation}
The background term defined above is to be compared with (7.38) of
\cite{DeFelice:2010aj} which is its analogue for the metric formalism
case. The crucial difference being that in our case the term is
proportional to $\dot\phi$ and vanishes in the absence of a scalar
field. 

The second-order differential equation can then be re-written in a
useful form by defining suitable Mukhanov-Sasaki variables \cite{Sasaki,Mukhanov:1990me},
$z_s:=a\sqrt{Q}$ and $u_s:=z_s\psi$. This allows us to write
(\ref{eqpsi}) as the equation for an oscillator with a time dependent
frequency which depends only on background quantities
\begin{equation}
u_s''+\left(k^2-\frac{z_s''}{z_s}\right)u_s=0\,,
\label{equ}
\end{equation}
where a prime denotes differentiation with respect to the conformal
time $\eta:=\int\frac{dt}{a} \,$. In analogy with the standard and
metric cases \cite{Stewart:1993bc,Hwang:1996xh}, we define the
so--called slow--roll parameters
\begin{eqnarray}
\epsilon_1:=-\frac{\dot{H}}{H^2} \qquad \epsilon_2:=\frac{\ddot\phi}{H\dot\phi} \qquad \epsilon_3:=\frac{\dot{F}}{2HF}\,,
\label{eq:slow-roll params}
\end{eqnarray}
where the additional $\epsilon_3$ parametrises the shape of the $f(R)$
function in analogy with the relation of the first two parameters with
the shape of the inflaton potential. The background term can also be
re-written in terms of the new parameters as
\begin{equation}
Q=\frac{\omega\dot\phi^2}{H^2(1+\epsilon_3)^2} \,.
\end{equation}
If $\dot\epsilon_1=0$ we can write the conformal time as \cite{Stewart:1993bc}
\begin{equation}
\eta=-\frac{1}{aH(1-\epsilon_1)} \,.
\end{equation}
If in addition $\dot\epsilon_2=\dot\epsilon_3=0$ the time dependent
background contribution to the frequency term can be written as
\begin{eqnarray}
\frac{z_s''}{z_s} = \frac{\nu_s^2-1/4}{\eta^2} \,,
\end{eqnarray}
with
\begin{eqnarray}
\nu_s^2:=\frac{1}{4} + \frac{(1+\epsilon_1+\epsilon_2)(2+\epsilon_2)}{(1-\epsilon_1)^2} \,.
\end{eqnarray}
The solution of eq. (\ref{equ}) is then 
%given by a linear combination of Hankel functions as
%\begin{equation}
%u_s=\frac{1}{2} \sqrt{\pi|\eta|} e^{i\pi(1+2\nu_s)/4} \left[C_1H^{(1)}_{\nu_s}(k|\eta|) + C_2H^{(2)}_{\nu_s}(k|\eta|)\right] \,.
%\end{equation}
%
%In order to fix the constants of integration $C_1$ and $C_2$, we consider the asymptotic past ($k\eta\rightarrow-\infty$) solution of eq. (\ref{equ}). This is determined by a quantum field vacuum state and given by $u_s\rightarrow e^{-ik\eta}/\sqrt{2k}$, which implies $C_1=1$ and $C_2=0$. The solution we must take is then
%
\begin{equation}
u_s=\frac{1}{2} \sqrt{\pi|\eta|} e^{i\pi(1+2\nu_s)/4} H^{(1)}_{\nu_s}(k|\eta|) \,, \label{solus}
\end{equation}
which we have normalised by taking the early time, short wavelength
($k\eta \rightarrow \infty$) limit of the general solution which
reduces to the adiabatic vacuum case for a field in an expanding
background $u_s\rightarrow e^{-ik\eta}/\sqrt{2k}$. The power spectrum
of the curvature perturbation can then be defined as
\begin{eqnarray}
\mathcal{P}_s &:=& \frac{k^3}{2\pi^2} |\psi|^2 \nonumber\\
 &\simeq& \frac{1}{Q} \left((1-\epsilon_1)\frac{\Gamma(\nu_s)}{\Gamma(3/2)}\frac{H}{2\pi}\right)^2 \left(\frac{|k\eta|}{2}\right)^{3-2\nu_s} \,, \quad
\label{powerspectrum}
\end{eqnarray}
where we have taken the super-horizon limit ($k\eta\rightarrow 0$) of
the solution.
%
%Plugging in the solution (\ref{solus}), we can find the power spectrum in the super-horizon limit ($k\eta\rightarrow 0$)
%
%\begin{equation}
%\mathcal{P}_s \simeq \frac{1}{Q} \left((1-\epsilon_1)\frac{\Gamma(\nu_s)}{\Gamma(3/2)}\frac{H}{2\pi}\right)^2 \left(\frac{|k\eta|}{2}\right)^{3-2\nu_s} \,.
%\label{powerspectrum}
%\end{equation}
%

Since the scalar perturbation is conserved after exiting the Hubble
radius we can evaluate the power spectrum (\ref{powerspectrum}) at the
time when $k=aH$ and examine its spectral index $n_s$ defined as
\begin{equation}
n_s-1 := \left.\frac{d\ln\mathcal{P}_s}{d\ln k}\right|_{k=aH} = 3-2\nu_s\,.
\end{equation}
During inflation we have $\epsilon_i\ll 1$ ($i=1,2,3$), hence at first order in the $\epsilon$'s the scalar spectral index becomes
\begin{equation}
n_s-1 \simeq -4\epsilon_1-2\epsilon_2 \,.
\label{eq:scalspecindex}
\end{equation}
This result is identical to the standard case arising from the
Einstein-Hilbert action minimally coupled to an inflaton field in the
metric formalism. It is however distinct from the generalised gravity
case with metric formalism in that the scalar spectral index $n_s$ has
no explicit dependence, at linear order, on $\epsilon_3$, or rather,
the choice of function $f(R)$.

\subsection*{Tensor Perturbations}

A gravitational wave can be decomposed into its two polarization
states $h_\times$ and $h_+$ which both obey (\ref{gweq}). In the
following we will omit the subscripts and denote both states by $h$ as
they are decoupled and follow the same solutions. In defining their
power spectra we will take into account the presence of two complex
modes to which reality conditions apply. In analogy with the scalar
case (\ref{gweq}) can also be redefined as an equation for an
oscillator with time dependent frequency by introducing the
Mukhanov-Sasaki variable $u_t:=z_th$ with $z_t:=a\sqrt{F}$ such that
\begin{equation}
u_t''+\left(k^2-\frac{z_t''}{z_t}\right)u_t=0\,.
\end{equation}
Following the same route as the previous section we can define the
background contribution as 
%Again, if $\dot\epsilon_i=0$ ($i=1,2,3$) we can write
%
\begin{eqnarray}
\frac{z_t''}{z_t}=\frac{\nu_t^2-1/4}{\eta^2} \,,
\end{eqnarray}
with
\begin{eqnarray}
\nu_t^2 := \frac{1}{4} + \frac{(1+\epsilon_3)(2-\epsilon_1+\epsilon_3)}{(1-\epsilon_1)^2}\,,
\end{eqnarray}
and obtain the solution for the mode $u_t$
% the solution is still given in terms of Hankel functions
\begin{equation}
u_t=\frac{1}{2} \sqrt{\pi|\eta|} e^{i\pi(1+2\nu_t)/4} H^{(1)}_{\nu_s}(k|\eta|) \,.
\label{solut}
\end{equation}
%This time the contribution given by $f$ is not ruled out as $\epsilon_3$ appears explicitly in $\nu_t$.
The spectrum of tensor perturbations (taking into account polarization
states) is then defined as
\begin{eqnarray}
\mathcal{P}_t &:=&  \frac{4\,k^3}{\pi^2}|h|^2 \nonumber\\
&\simeq& \frac{8}{F} \left((1-\epsilon_1)\frac{\Gamma(\nu_t)}{\Gamma(3/2)}\frac{H}{2\pi}\right)^2 \left(\frac{|k\eta|}{2}\right)^{3-2\nu_t} \,, \quad
\end{eqnarray}
where we have once again taken the super-horizon limit
($k\eta\rightarrow 0$). Similarly, we can evaluate the tensor spectral
index as
\begin{equation}
n_t:= \left.\frac{d\ln\mathcal{P}_t}{d\ln k}\right|_{k=aH} = 3-2\nu_t\,,
\end{equation}
which during inflation ($\epsilon_i\ll1$) reduces to
\begin{equation}
n_t \simeq -2\, (\epsilon_1 +\epsilon_3) \,.
\label{ten spec index}
\end{equation}
%To recover the standard result given by GR plus a minimally coupled scalar field we set $F=\omega=1$, in which case the tensor spectral index becomes $n_t\simeq -2\epsilon_1$. 

Finally, we can evaluate the tensor-to-scalar ratio $r$, defined as the ratio between the tensor and scalar spectra respectively. We find
\begin{equation}
r:=\frac{\mathcal{P}_t}{\mathcal{P}_s} \simeq \frac{8Q}{F} \simeq 16\, (\epsilon_1+\epsilon_3)\,,
\label{sca-ten ratio}
\end{equation}
which also introduces the consistency relation between the tensor
spectral index and tensor-to-scalar ratio $r=-8 n_t$.
%
%where $Q$ is given in eq. (\ref{defQ}). In this case the ratio 
%In the standard inflation theory we simply have $r\simeq \frac{\pi}{2}\epsilon_1$ DA CONTROLLARE

\section{Primordial spectra from inflation in specific classes of theories}
\label{sec:examples}

We now apply the results derived in the previous section to three
classes of theories which have been given particular attention in
the literature; $f(R)$ theories, Scalar-Tensor theories
and non-minimally coupled inflation.

\subsection*{$f(R)$ Theories}

We first look at $f(R)$ theories where the acceleration of the
inflationary expansion is driven solely by the $f(R)$ modifications to the
action as opposed to a potential dominated, minimally coupled
scalar. In attempting to define such a model in the Palatini formalism
an important constraint arises; whereas $f(R)$ theories in the metric
formalism introduce an extra dynamical, scalar degree of freedom, the
same theory in the Palatini formalism introduces only a non-dynamical
scalar mode with an algebraic constraint. This is most clearly seen by
relating both cases to Brans-Dicke theories with an explicit scalar
degree of freedom \cite{Sotiriou:2008rp} and deriving the scalar
equation of motion by varying the action obtained. In the Palatini
case the second-order terms in the scalar cancel out exactly resulting
in a non-dynamical scalar mode, although this is not obvious when first
inspecting the action.

This result is evident in our treatment in that the second-order term
in (\ref{eqpsi}) is proportional to $Q$ which is not defined
in the absence of the scalar mode $\phi$. Indeed in this case the
equation reduces to a first-order constraint
\begin{equation}
  \left( 3H + \frac{3\dot F}{2F}
\right)\dot\psi + \frac{k^2}{a^2}\psi = 0\,.
\end{equation}
This shows that in the Palatini formulation of $f(R)$ theories any
initial curvature perturbation decays during inflation and are not
useful for seeding the growth of structure. Thus Palatini inflation
driven solely by $f(R)$ modifications can only be a viable model with
the addition of an auxiliary slowly rolling scalar field. This
requirement negates the theoretical motivation behind this kind of
model and in addition our analysis does not account for this
possibility due to the restriction we made for imposing our choice of gauge.

\subsection*{Scalar-Tensor Theories}

We are then constrained to work with Scalar-Tensor theories in
obtaining a model where Palatini generalised gravity successfully
drives inflation. However, we can also derive some general results for
the observables in this case where, as previously noted,
$f(R,\phi)\rightarrow F(\phi)\,R$. In this case (\ref{acceq})
leads to a constraint between $\epsilon_1$ and $\epsilon_3$
\begin{equation}
  (\epsilon_1+\epsilon_3)(1+\epsilon_3)= \epsilon_3^2 \frac{2\,\omega\,F}{F^2_{,\phi}}\,,
\end{equation}
where no further $\epsilon_3$ contribution appears for general choices
of $\omega(\phi)$ and $F(\phi)$. To first order in $\epsilon_1$ and
$\epsilon_3$ this leads to the constraint
\begin{equation}
  \epsilon_1+\epsilon_3 \simeq 0\,.
\label{eq:no tensor in st}
\end{equation}
This has an immediate impact on the observables since both the tensor
spectral index $n_t$ and scalar-to-tensor ratio $r$ are proportional
to $\epsilon_1+\epsilon_3$ (see (\ref{ten spec index}) and
(\ref{sca-ten ratio})). Thus although the Palatini formulation of
scalar-tensor theories could drive inflation and produce nearly scale
invariant scalar perturbations, in agreement with observations, they
will not produce any tensor modes.  This is a general features of
Palatini Scalar-Tensor theories, which consequently might have to be
discarded as possible models for inflation if even a small amount
of tensor modes are observed by future CMB experiments.

\subsection*{Non-Minimally Coupled Inflation}

As an example of Scalar-Tensor gravity we analyse the case of
non-minimally coupled inflation \cite{Salopek:1988qh} which has been
recently used as an attempt to identify the inflaton field with the
standard model Higgs boson \cite{Bezrukov:2007ep}. In these models we
have $F(\phi)=1+\xi\phi^2$, $\omega(\phi)=1$ and
$V(\phi)=\lambda(\phi^2-v^2)^2$, where $\xi$, $\lambda$, $v$ are
parameters of the theory. Slow-Roll inflation is obtained in the
region $\xi\phi^2\gg v^2$, where the potential becomes
$V(\phi)\simeq\lambda\phi^4$. The slow-roll parameters
(\ref{eq:slow-roll params}) at first order in the slow-roll conditions
are
\begin{eqnarray}
\epsilon_1&\simeq&-4\xi\,\frac{\xi\phi^2-1}{\xi\phi^2+1}\,, \\ \epsilon_2&\simeq& 4\xi\,\frac{(\xi\phi^2+3)(2\xi\phi^2-1)}{\xi\phi^2(\xi\phi^2+1)}\,, \\ \epsilon_3&\simeq& 4\xi\,\frac{\xi\phi^2-1}{\xi\phi^2+1}\,.
\end{eqnarray}
According to the general result we obtained above for Scalar-Tensor
theories (\ref{eq:no tensor in st}), we immediately note that
$\epsilon_1$ is exactly the opposite of $\epsilon_3$. As expected then
the tensor spectral index and tensor-to-scalar ratio vanish at first
order. Using (\ref{eq:scalspecindex}) the scalar spectral index can be
written as
\begin{equation}
n_s-1\simeq \frac{8}{\phi^2}\frac{3-7\xi\phi^2}{1+\xi\phi^2} \,.
\end{equation}
In order to constrain $\xi$ with this result we need a value for
$\phi$ to substitute in. This can be obtained from the minimum number
of $e$-foldings $N$ which occurred during inflation,
\begin{eqnarray}
N=\int^{\phi_{\rm end}}_{\phi_{\rm start}}\frac{H}{\dot\phi}d\phi
\simeq -\frac{1}{8\xi}\log(1-\xi\phi^2_{\rm start})\,,
\label{eq:=e-foldings}
\end{eqnarray}
where $\phi_{\rm start}$ and $\phi_{\rm end}$ are the values of the inflaton
field at the beginning and end of inflation respectively and we have used
 the approximation $\phi_{\rm start}\gg\phi_{\rm end}$  in the last
step. Considering $N\simeq 62$ and an observed scalar spectral
index of $n_s \simeq 0.96$, we obtain the (positive) value for $\xi$ of
\begin{eqnarray}
\xi \simeq 2.9\cdot 10^{-3}\,.
%\xi \simeq -4.2\cdot 10^{-3} \,.
\end{eqnarray}
The value is small, which means small deviation from minimal
coupling. This is in agreement with the metric case result found in
\cite{Nozari:2010uu}. However, non-minimally coupled inflationary
models in the regime $\xi\gg 1$ are also considered viable
\cite{Bezrukov:2007ep} (see also \cite{Bauer:2008zj} for the Palatini
approach). The latter are indeed the models which try to unify the
inflaton with the Higgs boson \footnote{There is some confusion as to
  the observable-driven value of the $\xi$ parameter due to the use of
  both Jordan and Einstein frames in deriving observables (See
  discussion in \cite{Nozari:2010uu}). Large values of $\xi$, required
  to unify inflation and Higgs fields, have only been discussed in
  analysis based in the Einstein frame.}.

Finally, given the value for $\xi$ and imposing the COBE normalization
for the curvature perturbation power spectrum
\begin{equation}
\mathcal P_s \simeq \left.\left(\frac{H^2}{2\pi\dot\phi}\right)^2\right|_{aH=k} \simeq \frac{(0.027)^4}{24\pi^2}\,,
\end{equation}
we can also obtain a value of $\lambda$:
\begin{equation}
\lambda \simeq 2.3 \cdot 10^{-14}\,.
% \qquad \lambda \simeq -3.6 \cdot 10^{-13} \,.
\end{equation}
Again, this result seems to agree with the metric case
\cite{Nozari:2010uu}, where small values of $\lambda$ are also
obtained.

\section{Conclusions}\label{sec:disc}

We have analysed the generation of primordial scalar and tensor
perturbations during inflation in the Palatini formulation of
generalised gravity. A general result, derived in this work, is that
the scalar spectral index does not depend on the choice of $f(R)$
function explicitly and, to first order in slow-roll parameters, we
recover the same expression for $n_s$ as the standard case. This is
distinct from the metric case where $n_s$ depends explicitly on the
form of $f(R)$ at the same order. On the other hand, we find that the
tensor mode spectral index $n_t$ and the tensor-to-scalar ratio $r$
depend on $\epsilon_3$, the generalised slow-roll parameter defined by
the choice of $f(R)$. This result is the same as the metric case.

In restricting the analysis to specific classes of theories we also
find that models of inflation driven purely by $f(R)$ modifications
are not viable scenarios in the Palatini framework for generating
primordial curvature perturbations that seed structure formation. This
is specific to the Palatini treatment and is in contrast to the metric
formalism where the extra scalar degree of freedom, dynamical in this
case, allows the evolution of perturbations. 

The other interesting class of theories, namely Scalar-Tensor models,
result in a specific prediction that tensor modes vanish at first
order in the slow-roll parameters. This prediction will soon be tested
by CMB experiments and can potentially falsify these models. As a
particular example we analyse the case of non-minimally coupled
inflation where we find that constraints on the scalar spectral index allow
only small values of the coupling parameters of the theory. Thus for
the case considered the theory is constrained to a regime which is not
of interest in effort to identify the inflaton field with the Higgs
boson.

It should be stressed that the classes of models analysed in this work
are restricted by our choice of gauge to either pure $f(R)$ or
Scalar-Tensor cases. More general choices of $f(R,\phi)$ depending on
both $R$ and $\phi$ have not been constrained here and may provide
viable models for early or late-time acceleration.

\end{document}